\documentclass[12pt]{article}
\usepackage{epsfig}
\usepackage{amssymb}
\usepackage{amsmath}
\usepackage{amsfonts}
\usepackage{graphicx}
\usepackage{mathrsfs}
\usepackage[dvips]{color}
\usepackage{multirow}

\newcommand{\fa}{\mathfrak{a}}
\newcommand{\fb}{\mathfrak{b}}

\newcommand{\fg}{\mathfrak{g}}

\newcommand{\fn}{{\,\mathfrak{n}\,}}

\newcommand{\fz}{\mathfrak{z}}

\newcommand{\bM}{\mathbf{M}}

\newcommand{\cP}{\mathcal{P}}

\newcommand{\cU}{\mathcal{U}}

\newcommand{\be}{\begin{equation}}
\newcommand{\ee}{\end{equation}}
\newcommand{\bea}{\begin{eqnarray}}
\newcommand{\eea}{\end{eqnarray}}
\newcommand{\nn}{\nonumber}

\newcommand{\ed}{\end{document}}

\newcommand{\bi}{\begin{itemize}}
\newcommand{\ei}{\end{itemize}}

\newcommand{\bce}{\begin{center}}
\newcommand{\ece}{\end{center}}

\newcommand{\sH}{\mathscr{H}}

\newcommand{\sU}{\mathscr{U}}
\newcommand{\RE}{{\rm Re}}
\newcommand{\IM}{{\rm Im}}

\begin{document}

\title{A Dynamical Formulation of One-Dimensional Scattering Theory and Its Applications in Optics}

\author{Ali~Mostafazadeh\thanks{E-mail address:
amostafazadeh@ku.edu.tr, Phone: +90 212 338 1462, Fax: +90 212 338
1559}
\\
Department of Mathematics, Ko\c{c} University,\\
34450 Sar{\i}yer, Istanbul, Turkey}
\date{ }
\maketitle

\begin{abstract} We develop a dynamical formulation of one-dimensional scattering theory where the reflection and transmission amplitudes for a general, possibly complex and energy-dependent, scattering potential are given as solutions of a set of dynamical equations. By decoupling and partially integrating these equations, we reduce the scattering problem to a second order linear differential equation with universal initial conditions that is equivalent to an initial-value time-independent Schr\"odinger equation. We give explicit formulas for the reflection and transmission amplitudes in terms of the solution of either of these equations and employ them to outline an inverse-scattering method for constructing finite-range potentials with desirable scattering properties at any prescribed wavelength. In particular, we construct optical potentials displaying threshold lasing, anti-lasing, and unidirectional invisibility.
\medskip

\end{abstract}

\maketitle

\section{Introduction}

Consider a possibly complex and energy-dependent scattering potential $v(x)$ that satisfies the vanishing boundary condition at infinity, i.e., $v(x)\to 0$ as $x\to\pm\infty$. The general solution of the time-independent Schr\"odinger equation,
	\be
	-\psi''(x)+v(x)\psi=k^2\psi(x),
	\label{sch}
	\ee
has the following asymptotic form.
	\be
	\psi(x)=A_\pm e^{ikx}+B_\pm e^{-ikx}~~~{\rm for}~~~x\to\pm\infty,
	\label{asym}
	\ee
where $A_\pm$ and $B_\pm$ are complex coefficients. All the scattering properties of the potential are encoded in the associated transfer matrix $\bM$ which is defined by the condition:
	\be
	\left[\begin{array}{c}
	A_+\\ B_+\end{array}\right]=\bM\left[\begin{array}{c}
	A_-\\ B_-\end{array}\right].
	\label{M}
	\ee
The entries $M_{ij}$ of $\bM$ are related to the left/right reflection and transmission amplitudes, $R^{r/l}$ and $T$, according to \cite{prl-2009,note1}
	\be
	\begin{aligned}
	&M_{11}=T- R^l R^r/T,~~~~~~ && M_{12}=R^r/T,\\
	&M_{21}=- R^l/T, && M_{22}=1/T.
	\end{aligned}
	\label{M-RT}
	\ee

The transfer matrix has a number of interesting properties. It has a unit determinant. The zeros of $M_{22}$ in the complex $k^2$-plane correspond to the bound states, resonances, and anti-resonances of $v$. In particular, the real zeros yield the zero-width resonances that are called spectral singularities \cite{prl-2009,prl-2013}. In optics, these give rise to threshold lasing \cite{pra-2011a}. The real zeros of $M_{11}$ are the energy values at which the potential acts as a coherent perfect absorber (CPA) \cite{longhi-2010}. This is also known as an anti-laser \cite{anti-laser}. The real zeros of $M_{12}$ and $M_{21}$ are respectively the energies at which $v$ is reflectionless from the right and the left. If at such an energy only one of $M_{12}$ and $M_{21}$ vanishes while $M_{22}=1$, the system displays unidirectional invisibility \cite{pra-2013a}. The latter has important applications in devising unidirectional optical devices, and has been a subject of extensive theoretical \cite{invisible-0,invisible,invisible-other,pra-2013a} and experimental studies \cite{exp} in the past two years. The problem of constructing optical potentials with any of the above-mentioned properties is therefore of utmost importance. In this article we develop a formulation of scattering theory that, besides its conceptual and practical advantages, offers a simple solution for this kind of inverse scattering problems.

An important property of the transfer matrix is its composition property \cite{sanchez}: Consider the truncated potential
	\be
	v_a(x):=v(x)\theta(a-x)=
    \left\{\begin{array}{ccc}
    v(x) &{\rm for} & x\leq a,\\
    0 &{\rm for} & x>a,\end{array}\right.
	\label{truncate}
	\ee
where $a$ is a real number and $\theta(x)$ is the step function with values $0$ and $1$ respectively for $x<0$ and $x\geq 0$. Let $\bM_1$ and $\bM_2$ be respectively the transfer matrix for $v_a$ and $v-v_a$. Then, $\bM_2\bM_1=\bM$. This relation plays a particularly useful role in modeling and the numerical investigation of various physical phenomena in optics \cite{optics}, condensed matter physics \cite{CMP}, and acoustics \cite{Accu}.

The composition property of the transfer matrix is shared by another quantum mechanical quantity of central importance, namely the evolution operator $U(t,t_0)$ of any Hamiltonian operator; if we denote $U(t,t_0)$, $U(t_1,t_0)$, and $U(t,t_1)$ respectively by $U$, $U_1$, and $U_2$, with $t_1\in[t,t_0]$, we have $U_2U_1=U$. This simple observation leads to the natural question whether we can relate $\bM$ to the evolution operator of a $2\times 2$ matrix Hamiltonian. As we show below the answer to this question is in the affirmative.

\section{Transfer Matrix Given by an Evolution Operator}

Consider the two-component state vector:
	\be
	\Psi:=\frac{1}{2}\left[\begin{array}{c}
	e^{-ikx}(\psi-ik^{-1} \psi' )\\[3pt]
	 e^{ikx}(\psi+ik^{-1} \psi' )\end{array}\right].
	\ee
It is easy to show that if we express $\Psi$ as a function of $\tau:=kx$, then $\psi$ is a solution of (\ref{sch}) if and only if $\Psi$ satisfies the Schr\"{o}dinger equation, $i\frac{d}{d\tau}\Psi(\tau)=\sH(\tau)\Psi(\tau)$, for the following singular, traceless, non-Hermitian, non-diagonalizable, pseudo-normal \cite{note2} matrix Hamiltonian.
	\bea
	 \sH(\tau)&:=&\frac{v(\tau/k)}{2k^2}\left[\begin{array}{cc}
	1 & e^{-2i\tau}\\
	-e^{2i\tau}&-1\end{array}\right].
	\label{sH=}
	\eea
Furthermore, because (\ref{asym}) corresponds to
    \[\Psi(\tau)\to\left[\begin{array}{c} A_\pm\\ B_\pm\end{array}\right]~~{\rm as}~~\tau\to\pm\infty,\]
Eq.~(\ref{M}) implies
	\be
	 \bM=\lim_{\tau_\pm\to\pm\infty}\sU(\tau_+,\tau_-)=:\sU(\infty,-\infty),
	\label{M=U}
	\ee
where $\sU(\tau,\tau_0)$ is the evolution operator associated with $\sH(\tau)$. This is the matrix-valued function of the real variables $\tau$ and $\tau_0$ that fulfills
	\be
	 i\frac{d}{d\tau}\sU(\tau,\tau_0)=\sH(\tau)\sU(\tau,\tau_0),~~~~\sU(\tau_0,\tau_0)=1.
	\label{sch-eq}
	\ee

The mathematical properties of $\bM$ can be easily derived from Eq.~(\ref{M=U}). For example, because $\sH(\tau)$ is traceless, $\det\sU(\tau,\tau_0)=1$. This relation together with (\ref{M=U}) imply $\det\bM=1$. Similarly, we can establish the composition property of $\bM$ from that of $\sU(\tau,\tau_0)$. This requires considering the transfer matrix of the truncated potential (\ref{truncate}) that we denote by $\bM(\alpha)$ with $\alpha:=ka$.

An immediate consequence of Eqs.~(\ref{truncate}), (\ref{sH=}), and (\ref{M=U}) is the fact that $\bM(\alpha)=\lim_{\alpha_-\to-\infty}\cU(\alpha,\alpha_-)=:\cU(\alpha,-\infty)$. According to (\ref{sch-eq}), this implies
	\be
	 i\frac{d}{d\alpha}\bM(\alpha)=\sH(\alpha)\bM(\alpha).
	\label{dyn-eq}
	\ee
We also have
	\bea
	 &&\bM(-\infty):=\lim_{\alpha\to-\infty}\bM(\alpha)=1,
	\label{ini}\\
	 &&\bM=\lim_{\alpha\to\infty}\bM(\alpha)=:\bM(\infty).
	\label{M=M}
	\eea
Because $\bM(\alpha)$ is invertible, we can express (\ref{dyn-eq}) as
	\be
	 i\left[\frac{d}{d\alpha}\bM(\alpha)\right]\bM(\alpha)^{-1}=\sH(\alpha).
	\label{dyn-eqn}
	\ee
In view of the particular structure of $\sH(\alpha)$, this equation restricts the form of $\bM(\alpha)$ enormously. Specifically, it requires the entries of the matrix $\left[\frac{d}{d\alpha}\bM(\alpha)\right]\bM(\alpha)^{-1}$, that we denote by $N_{ij}$, to satisfy
	 \[N_{11}=e^{2i\alpha}N_{12}=-e^{-2i\alpha}N_{21}=-N_{22}=-\frac{iv(\alpha/k)}{2k^2}.\]

As a nontrivial check on our calculations, consider verifying (\ref{dyn-eqn}) for a rectangular barrier potential of a possibly complex hight $\fz$, i.e.,
	\be
	v(x)=\fz\,\theta(x)\theta(L-x),
	\label{barrier}
	\ee
where $L$ is a positive number, \cite{pra-2011a}.  The form of the transfer matrix $\bM(\alpha)$ for this potential is well-known \cite{pra-2013a}. $\bM(\alpha)$ takes the values $1$ and $\bM(kL)$ respectively for $\alpha<0$ and $\alpha\geq kL$. Its entries have the following form for $\alpha\in[0,kL]$.
	{\begin{align}
	 &M_{11}(\alpha)=\left[\cos(\!\fn\alpha)+i(\fn^2+1)\sin(\!\fn\alpha)/2\fn\right]e^{-i\alpha},
	\label{M11=}\\
	&M_{12}(\alpha)=i(\fn^2-1)\sin(\!\fn\alpha) e^{-i\alpha}/2\fn,
	\label{M12=}\\
	 &M_{21}(\alpha)=M_{12}(-\alpha),~~~~~M_{22}(\alpha)=M_{11}(-\alpha),
	\label{M21M22=}
	\end{align}}%
where $\fn:=\sqrt{1-\fz/k^2}$. Substituting these relations in the left-hand side of (\ref{dyn-eqn}), we find, after miraculous simplifications, that it really gives the expression (\ref{sH=}) for $\sH(\alpha)$ with $v(\alpha/k)=\fz$.

\section{Dynamical Equations Of 1-D Scattering Theory}
	
Equation (\ref{dyn-eqn}) together with (\ref{ini}) and (\ref{M=M}) provide a dynamical formulation of one-dimensional scattering theory \cite{muga}. In view of (\ref{M-RT}), we can use them to obtain dynamical equations for the reflection and transmission  amplitudes, $R^{r/l}(\alpha)$ and $T(\alpha)$, of the potential $v_a(x)$. We have decoupled and partially integrated these equations to express $R^l(\alpha)$ and $T(\alpha)$ in terms of $R^r(\alpha)$. The latter satisfies a Ricatti equation provided that we make the change of variable:
    \[\alpha\to z:=e^{-2i\alpha}=e^{-2ika}.\]
For a finite-range potential that vanishes outside an interval of the form $[a_-,a_+]$, we have
	\begin{align}
	&\frac{d}{dz}R^r(z)-\frac{\check v(z)}{4k^2z^2}\left[ R^r(z) +z\right]^2=0,
	\label{e1}\\
	&R^l(z)=
	\int_{z_-}^z dw\:S(w)^{-2}\frac{d}{dw}R^r(w),
	\label{e2}\\
	&T(z)=\left[R^r(z)+z\right]S(z)^{-1},
	\label{e3}
	\end{align}
where $\check v(z):=v(i\ln z/(2k))=v(x)$, $z_-:=e^{-2ika_-}$,
	\be
    S(z):=z_-\exp\left[\int_{z-}^z \frac{dw}{R^r(w)+w}\right],
    \label{S=9}
    \ee
and integrals in (\ref{e2}) and (\ref{S=9}) are to be evaluated along the circular arc:  $\{w=e^{-2i\varphi}|\varphi\in[\alpha_-,\alpha]\}$ in the complex plane. Notice that in the derivation of (\ref{e2}) and (\ref{e3}) we have imposed the initial conditions: $R^l(z_-)=0$ and $T(z_-)=1$, and that Eq.~(\ref{e1}) is to be solved together with the initial condition: $R^r(z_-)=0$.

Next, we use Eqs.~(\ref{e2}) -- (\ref{S=9}) to establish
	\bea
	R^r(z)&=&\frac{S(z)}{S'(z)}-z,
	\label{e5}\\
	R^l(z)&=&-\int_{z_-}^z dw~\frac{S''(w)}{S(w)S'(w)^2},
	\label{e2nn}\\
	T(z)&=&\frac{1}{S'(z)}.
	\label{e3nn}
	\eea
With the help of (\ref{e5}), we can show that (\ref{e1}) together with $R^r(z_-)=0$ is equivalent to the initial-value problem:
	\bea
	&&z^2S''(z)+\left[\frac{\check v(z)}{4k^2}\right]S(z)=0,
    \label{e1n}\\
	&&S(z_-)=z_-,~~~~~S'(z_-)=1.
    \label{ini-condi}
	\eea
It is remarkable that in terms of $\psi_{k-}(a):=e^{ika}S(e^{-2ik a})$, (\ref{e1n}) takes the form of the time-independent Schr\"odinger equation~(\ref{sch}),
    \be
	-\psi_{k-}''(a)+\left[v(a)-k^2\right]\,\psi_{k-}(a)=0,
    \label{e1n1}
    \ee
while (\ref{ini-condi}) becomes
    \be
    \psi_{k-}(a_-)=e^{-ika_-},~~~~~~\psi_{k-}'(a_-)=-ik e^{-ika_-}.
	\label{e1n2}
    \ee
These relations identify $\psi_{k-}$ with the Jost solution of (\ref{sch}) that is defined by the asymptotic boundary condition: $\psi_{k-}(x)\to e^{-ikx}$ as $x\to-\infty$, \cite{prl-2013}.

We can express the right-hand sides of (\ref{e5}) -- (\ref{e3nn}) in terms of $\psi_{k-}$. With the help of (\ref{e1n1}), this leads to the following expressions for the reflection and transmission amplitudes of the truncated potentials $v_a(x)$.
    \bea
	R^r(a)&=&-\frac{e^{-2ika} F_+(k,a)}{F_-(k,a)},
	\label{e21}\\
	R^l(a)&=&2ik\int_{a_-}^a dx~\frac{v(x)}{F_-(k,x)^2},
	\label{e22}\\
	T(a)&=&\frac{-2ik e^{-ika}}{F_-(k,a)},
	\label{e23}
	\eea
where $F_\pm(k,x):=\psi_{k-}'(x)\pm ik\psi_{k-}(x)$. Substituting $a_+$ for $a$ in these relations gives the reflection and transmission amplitudes, $R^{r/l}$ and $T$, for the original potential $v(x)$;
	\be
	 R^{r/l}=R^{r/l}(a_+),~~~~~~~T=T(a_+).
	\label{e4}
	\ee
For the infinite-range potentials we need to evaluate the $a_\pm\to\pm\infty$ limit of the right-hand side of these equations.

Equations~(\ref{e21}) and (\ref{e23}) agree with the formulas given in Ref.~\cite{prl-2013} for $R^r$ and $T$ in terms of the Jost solution~$\psi_{k-}$. These equations together with (\ref{e22}) reduce the scattering problem to the determination of the Jost solution~$\psi_{k-}$ of the time-independent Schr\"odinger equation (\ref{sch}). If we apply the above formulas to determine the reflection and transmission amplitudes of the parity-transformed potential, $v^{\cP}(x):=v(-x)$, and recall that under the parity transformation the transmission coefficient is left invariant while the left/right reflection amplitude are mapped to right/left reflection amplitude and the Jost solution $\psi_{k-}$ transforms to the Jost solution $\psi_{k+}$ given by $\psi_{k+}(x)\to e^{ikx}$ as $x\to\infty$, we can express $R^{r/l}$ and $T$ in terms of $\psi_{k+}$. Note however that in this way we find an expression for $R^r$ that is similar to (\ref{e22}) while the expression for $R^l$ does not involve an integration.

In practice, given a potential $v(x)$, we can solve the initial-value problem defined by (\ref{e1n1}) and (\ref{e1n2}) numerically and substitute the result in (\ref{e21}) -- (\ref{e23}) to determine the corresponding reflection and transmission amplitudes.

The treatment of the scattering problem that is based on the solution of the initial-value problem (\ref{e1n}) -- (\ref{ini-condi}) and Eqs.~(\ref{e5}) -- (\ref{e3nn}) is also useful. For example, it is easy to see that, for every pair of positive real numbers $k_0$ and $L$, Eq.~(\ref{e1n}) is exactly solvable for the potential
	\be
	v(x)=\fz\,e^{-4ik_0 x}\:\theta(x)\theta(L-x),
	\label{v=2}
	\ee
and the wavenumber $k=k_0$. Substituting the solution of (\ref{e1n}) -- (\ref{ini-condi}) in (\ref{e5}) -- (\ref{e3nn}) yields the following expressions for the reflection and transmission amplitudes.
	\begin{align}
	&R^r=(1+\fa \tan \fb)^{-1}-[\fa(\fa+\cot \fb)]^{-1}+\fa^{-1}\fb-1,\nn\\
	&R^l=-\fa(\cot\fb+\fa)^{-1},~~~~~~
	T=(\cos\fb+\fa\sin\fb)^{-1},\nn
	\end{align}
where $\fa:=\sqrt\fz/2k_0$ and $\fb:=\fa(1-e^{-2ik_0L})$. Notice that $R^{r/l}$ and $T$ are locally period functions of $k_0L$ with period $\pi$, while the potential (\ref{v=2}) is a locally period function of $k_0x$ with period $\pi/2$.

Eq.~(\ref{e1n}) is an exactly solvable (Euler) equation for the barrier potential (\ref{barrier}). Substituting the solution of this equation fulfilling the initial conditions (\ref{ini-condi}) in (\ref{e5}) -- (\ref{e3nn}) and using (\ref{e4}) and (\ref{M-RT}) give  (\ref{M11=}) -- (\ref{M21M22=}).

\section{Finite-Range Optical Potentials and Inverse Scattering}

In this section we confine our attention to finite-range potentials which are of particular interest in effectively one-dimensional optical systems. In particular, we consider situations where the optically active components are confined to an interval of the form $[0,L]$, i.e., set $a_-=0$ and $a_+=L$, so that $z_-:=e^{-2ika_+}=1$ and $z_+:=e^{-2ika_+}=e^{-2ikL}$.

A practically important feature of our formalism is that it allows for the construction of scattering potentials with appealing scattering properties at a given wavenumber $k_0$, i.e., it gives rise to a certain inverse scattering prescription that avoids the mathematical difficulties of the standard inverse scattering theory \cite{IST}. In the following we describe this prescription by trying to construct optical potentials supporting spectral singularities (threshold lasing), antilasing, and unidirectional invisibility.

We first recall that the Helmholtz equation for an effectively one-dimensional optical system has the form of the Sch\"odinger equation (\ref{sch}) with the potential
	\be
	v(x)=k^2[1-\fn(x)^2],
	\label{v-n}
	\ee
where $\fn(x)$ is the complex refractive index of the active component of the system, that we take to be confined in the interval $[0,L]$.

Suppose that we wish to determine the form of $\fn(x)$ such that the system acts as a threshold laser at a wavenumber $k_0$. This means that the potential $v(x)$ should possess a spectral singularity \cite{pra-2011a} at $k_0$. This happens if the transmission amplitude has a pole at $k=k_0$, \cite{prl-2009}. In view of (\ref{e3nn}), we can easily realize this requirement by choosing the form of $S(z)$ such that $S'(z_+)=0$. This choice should additionally satisfy the initial conditions (\ref{ini-condi}), i.e.,
    \be
    S(1)=S'(1)=1.
    \label{ini-condi-2}
    \ee
Once we make such a choice we can determine $v(x)$ using (\ref{e1n}).

A simple choice with the above properties is $S'(z)=(z-z_+)/(1-z_+)$. Integrating this expression and enforcing $S(1)=1$ give
    \[S(z)=\frac{z^2-2z_+z+1}{2(1-z_+)}.\]
Substituting this relation in (\ref{e1n}) and using (\ref{v-n}), we find
	\be
	\fn\!(x)^2=1+\frac{8\,\theta(x)\theta(L-x)}{e^{4ik_0x}-2 e^{-2ik_0(L-x)}+1}.
	\nn
	\ee
This is a non-singular function provided that $k_0L$ is not an integer multiple $\pi/2$.  If $\fn(x)^2$ is given by the complex-conjugate of the right-hand side of this equation, the system acts as a CPA at $k=k_0$.

\begin{figure}
	\begin{center}
	\includegraphics[scale=.7]{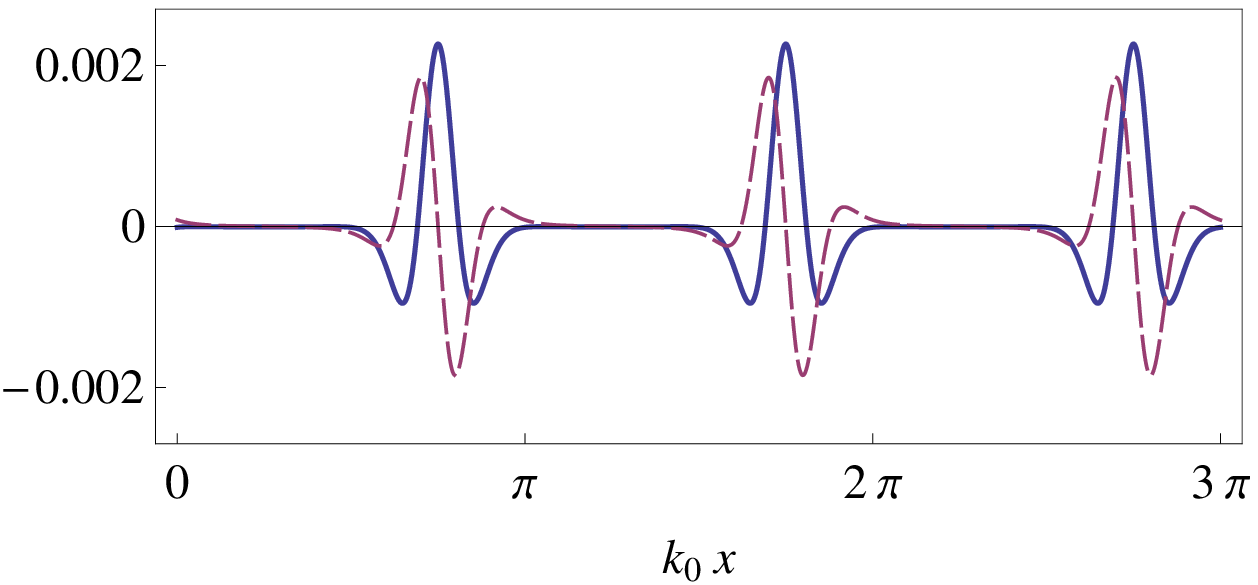}\\[12pt]
    \includegraphics[scale=.7]{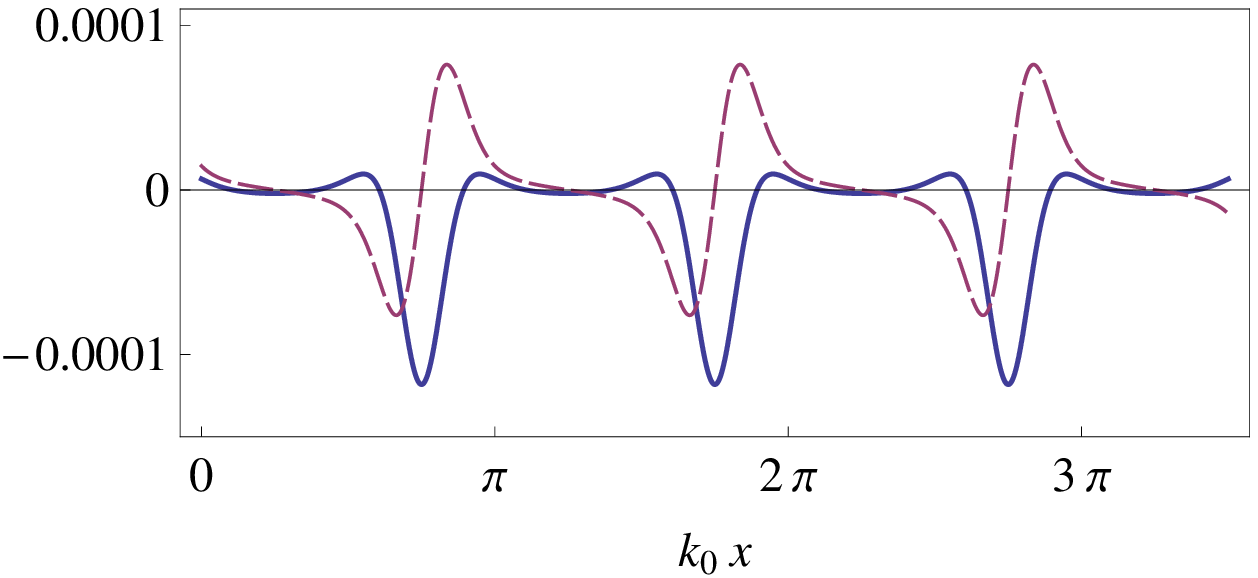}
	\caption{(Color online) Graph of $\RE(\fn^{\!\!2})-1$ (solid, blue curve) and $\IM(\fn^{\!\!2})$ (dashed, purple curve) as functions of $k_0x$ for $\fn^{\!\!2}$ given by (\ref{n=6}) with $\gamma=10^{-6}$ and $k_0L=3\pi$ (top curve) and $k_0L=7\pi/2$ (bottom curve).}
	\label{fig1}
	\end{center}
	\end{figure}

Next, we construct an optical potential that is unidirectionally reflectionless from the right for a wavenumber $k_0$, i.e., $R^r=0$ and $R^l\neq 0$ for $k=k_0$. In view of (\ref{e5}), we can ensure this condition by choosing the form of $S(z)$ such that $z_+S'(z_+)=S(z_+)$. A simple example that also satisfies (\ref{ini-condi-2}) is
	\be
	S(z)=\gamma[\fg_1(z-1)^3-\fg_2(z-1)^2] + z,
	\label{S=7}
	\ee
where $\gamma$ is a possibly complex free parameter and
    \[\fg_1:=z_++1,~~~~~~~~~ \fg_2:=(z_+- 1)(2z_++1).\]
Substituting (\ref{S=7}) in (\ref{e1n}) and using (\ref{v-n}), we find
    { \be
	\fn(x)^2=1 + \frac{8\gamma[3\fg_1(e^{-2ik_0x}-1)-\fg_2]
    \theta(x)\theta(L-x)}{e^{2ik_0x}+\gamma(1-e^{2ik_0x})^2(\fg_1\,e^{-2ik_0x}-\fg_1-\fg_2)}.
	\label{n=6}
	\ee}%
Here $\gamma$ controls the strength of the potential $v(x)$. For typical optical setups it takes very small real values. Eq.~(\ref{n=6}) takes a simpler form whenever $k_0L$ is an integer multiple $\pi$. In this case, $\fg_1=2$ and $\fg_2=0$, and we have
    \bea
	\fn\!(x)^2&=&1 + \frac{48\gamma(1-e^{2ik_0x})\,\theta(x)\theta(L-x)}{e^{4ik_0x}+2\gamma\,
(1-e^{2ik_0x})^3}.
	\label{n=61}
	\eea
Figure~\ref{fig1} shows the graphs of the real and imaginary parts of $\fn^2-1$ for $\gamma=10^{-6}$ and $k_0L=3\pi,7\pi/2$. A graphical examination of real and imaginary parts of $\fn^2-1$ shows that they change with the largest (respectively smallest) amplitude whenever $k_0L$ (respectively $k_0L-\pi/2$) is an integer multiple of $\pi$.

\begin{figure}
	\begin{center}
	\includegraphics[scale=.7]{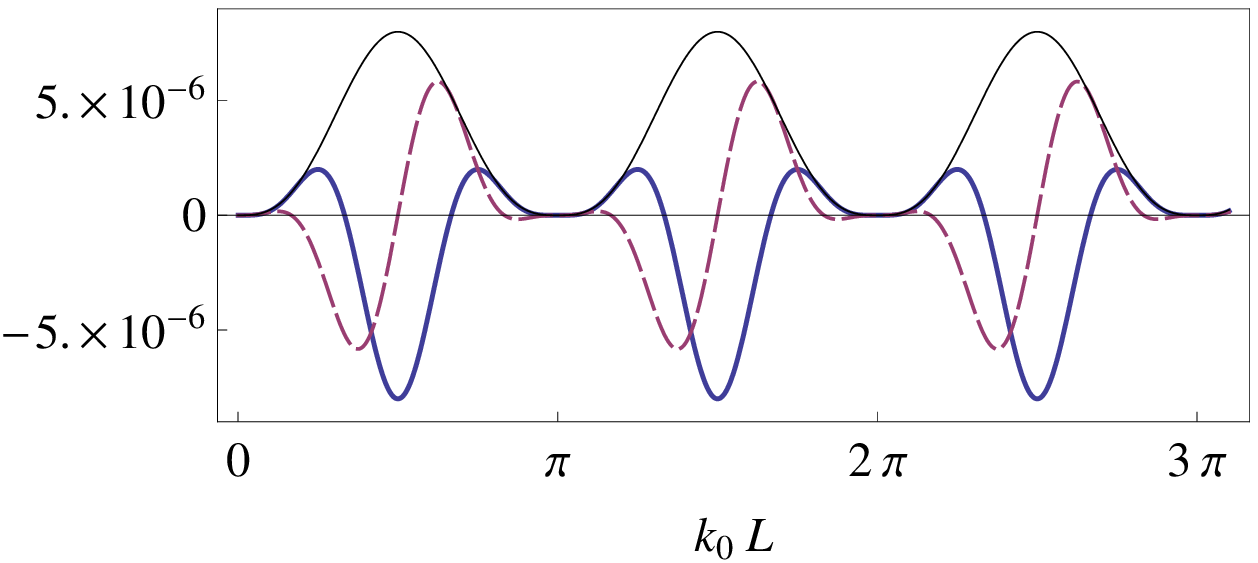}\\[12pt]
	\includegraphics[scale=.7]{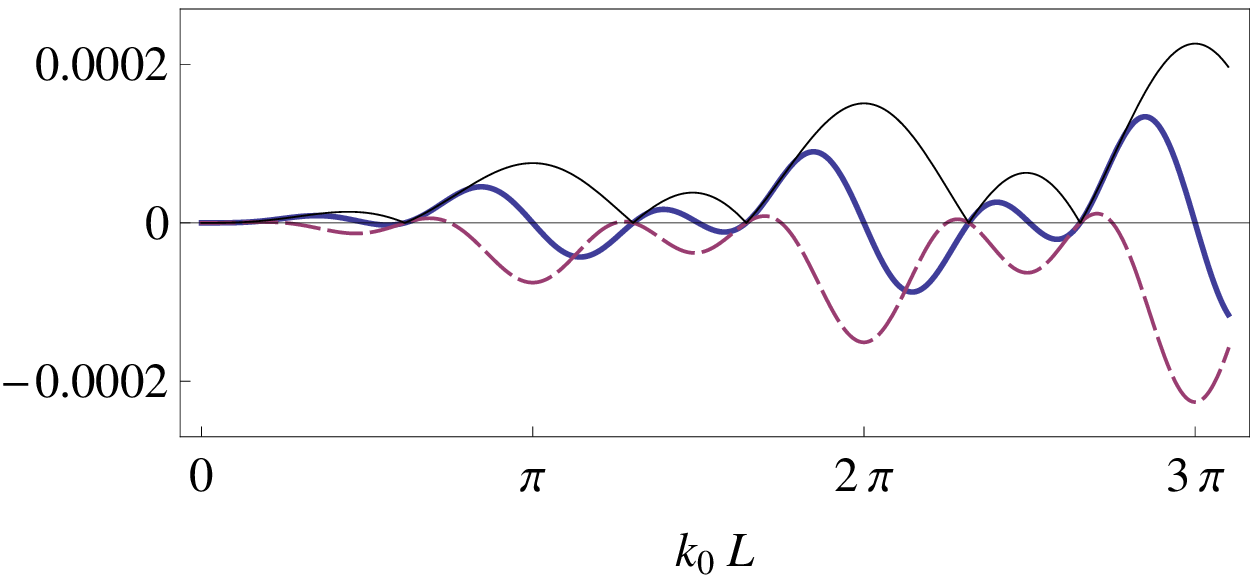}
	\caption{(Color online) Graph of the real part (solid, blue curve), imaginary part (dashed, purple curve) and absolute value (thin, black curve) of $T-1$ (top curve) and $R^l$ (bottom curve) as functions of $k_0L$ for $\fn^{\!\!2}$ give by (\ref{n=6}) with $\gamma=10^{-6}$.}
	\label{fig2}
	\end{center}
	\end{figure}

We can easily determine the transmission amplitude for (\ref{n=6}) by substituting (\ref{S=7}) in (\ref{e3nn}). This yields
	\bea
	T=\frac{1}{1+\gamma(1-e^{-2 i k_0 L})^3}.
	\label{T=6}
	\eea
In particular, $T$ is a locally periodic function of $k_0L$ with period $\pi$. The calculation of $R^l$ is much more involved. This is because the integrand in (\ref{e2nn}) is a sixth order rational function. For typical real values of $\gamma$, numerical evaluation of the right-hand side of (\ref{e2nn}) shows that $R^l$ is generically nonzero. For the cases where $k_0L$ is an integer multiple of $\pi$, the integral in (\ref{e2nn}) is a complex contour integral and we can use the residue theorem to study its structure. In this way, we could show that $R^l\neq 0$, while according to (\ref{T=6}), $T=1$. Therefore, the system is unidirectionally invisible from the right whenever $k_0L$ is an integer multiple of $\pi$. Figure~\ref{fig2} shows the graphs of the real part, imaginary part, and absolute value of $T-1$ and $R^l$ as a function of $k_0L$ for $\gamma=10^{-6}$.
As seen from this figure, $|R^l|$ attains its maxima at $k_0L=m\pi$ for positive integer values of $m$. Moreover, the hight of the maximum values of $|R^l|$ is a linearly increasing function of $m$. This is easy to justify, because for $k_0L=\pi m$, $R^l$ is given by a contour integral over a contour that encircles the pole(s) of the integrant $m$ times. This implies that $R^\ell=m R^l_1$, where $R^l_1$ is the left-reflection amplitude for $k_0L=\pi$. This observation shows that we can increase the value of $|R^l|$ using larger samples. A crude calculation shows that $|R^l|$ can take values as large as $1$, if we take $k_0L\gtrapprox 14300\pi$. For $\lambda_0:=2\pi/k_0\approx 1~\mu{\rm m}$, this corresponds to $L\gtrapprox 7.1~{mm}$. Note that this is achieved while $R^r=0$, $|\RE(T)-1|<10^{-5}$, and $|\IM(T)|<10^{-5}$. Therefore, the system is effectively unidirectionally invisible from the right for almost all large values of $k_0L$.

The above examples make use of extremely simple choices for the function $S$. We may consider more complicated choices with plenty of free parameters to construct large classes of optical potentials displaying threshold lasing, antilasing, unidirectional reflectionlessness or invisibility at a prescribed wavenumber, $k_0$. Our method produces the reflection and transmission amplitudes for these potentials at $k=k_0$. In order to determine these quantities for $k\neq k_0$, we can substitute the expression for the obtained potential in (\ref{e1n}) and solve it together with (\ref{ini-condi}) for $k\neq k_0$. In this way we find  $S$ for all $k$ and can use it in (\ref{e5}) -- (\ref{e3nn}) to compute $R^{r/l}$ and $T$ as a function of $k$. Alternatively, we may proceed solving (\ref{e1n1}) and (\ref{e1n2}) to determine $\psi_{k-}$ and use (\ref{e21}) -- (\ref{e23}) to find $R^{r/l}$ and $T$ for arbitrary $k$. In general, this requires using well-known numerical or approximate methods of solving second-order linear differential equations.

\section{Concluding Remarks}

In this article we have outlined a dynamical formulation of one-dimensional scattering theory with interesting conceptual and practical implications. This formulation is based on the simple idea of identifying the transfer matrix of the truncated potential $v_a(x)$ with the time-evolution operator for a non-Hermitian matrix Hamiltonian. This leads to a set of dynamical equations for the reflection and transmission amplitudes of $v_a(x)$ that reduce to an initial-value problem for a Ricatti equation, a second order linear equation, or the time-independent Schr\"odinger equation. The solution of the latter coincides with the Jost solution $\psi_{k-}(x)$. We have given explicit formulas for the reflection and transmission amplitudes in terms of  $\psi_{k-}$ as well as the solution of Eqs.~(\ref{e1n}) and (\ref{ini-condi}). This provides a convenient method of solving the scattering problem for finite-range potentials.

Another remarkable outcome of our approach is that it offers an extremely simple solution for a class of inverse scattering problems with immediate applications in optics. Specifically, it allows for the construction of large classes of optical potentials with interesting scattering properties at a prescribed wavenumber, such as threshold lasing, antilasing, and unidirectional invisibility.\\

\noindent {\bf Note:} Ref.~\cite{muga-1991}, that was brought to my attention by the referee, outlines a dynamical formulation of the time-independent Schr\"odinger equation for real scattering potentials. It involves using the hydrodynamical (polar) representation of this equation to obtain an iterative scheme for solving it.

\subsection*{Acknowledgments} I would like to thank Hugh Jones for bring Refs.~\cite{invisible-0} to my attention. This work has been supported by  the Scientific and Technological Research Council of Turkey (T\"UB\.{I}TAK) in the framework of the project no: 112T951, and by the Turkish Academy of Sciences (T\"UBA).

\ed
\begin{thebibliography}{99}

\bibitem{prl-2009} A.~Mostafazadeh, Phys.\ Rev.\ Lett.~\textbf{102}, 220402
(2009).

\bibitem{note1} Recall that $R^{r/l}$ and $T$ are defined by the asymptotic expression for the scattering (Jost) solutions $\psi_k^{l/r}$ of (\ref{sch}), namely $\psi_k^l(x)=\left\{
	\begin{array}{cc}
	e^{ikx}+R^l e^{-ikx} & {\rm for}~ x\to-\infty\\
	T e^{ikx}& {\rm for}~ x\to\infty
	\end{array}\right.$, and $
	\psi_k^r(x)=\left\{ \begin{array}{cc}
	T e^{-ikx} & {\rm for}~ x\to-\infty\\
	e^{-ikx}+R^r e^{ikx}& {\rm for}~ x\to\infty
	\end{array}\right..$

\bibitem{prl-2013} A.~Mostafazadeh, Phys.\ Rev.\ Lett.~\textbf{110}, 260402
(2013).

\bibitem{pra-2011a} A.~Mostafazadeh, Phys.\ Rev.\ A \textbf{83}, 045801 (2011).

\bibitem{longhi-2010} S.~Longhi, Physics \textbf{3}, 61 (2010); Phys.\ Rev.\ A  \textbf{82}, 031801 (2010) and Phys.\ Rev.\ A  \textbf{83}, 055804 (2011).

\bibitem{anti-laser} Y.~D.~Chong, L.~Ge, H.~Cao, and A.~D.~Stone, Phys.\ Rev.\ Lett.\ {\bf 105}, 053901 (2010); W.~Wan, Y.~Chong, L.~Ge, H.~Noh, A.~D.~Stone, and H.~Cao, Science \textbf{331}, 889 (2011); Y.~D.~Chong, L.~Ge, and A.~D.~Stone, Phys.\ Rev.\ Lett.~\textbf{106}, 093902 (2011); L.~Ge, Y.~D.~Chong, S.~Rotter, H.~E.~T\"ureci, and A.~D.~Stone, Phys.\ Rev.\ A~\textbf{84}, 023820 (2011); S.~Longhi, Phys.\ Rev.\ Lett.~\textbf{107}, 033901 (2011).

\bibitem{pra-2013a} A.~Mostafazadeh, Phys.\ Rev.\ A {\bf 87}, 012103 (2013).

\bibitem{sanchez} L.\ L.\ S\'anchez-Soto, J.\ J.\ Monz\'ona, A.\ G.\ Barriuso, and J.\ F.\ Cari$\tilde{\rm n}$ena, Phys.\ Rep.\ {\bf 513}, 191 (2012).

\bibitem{optics} J.~B.~Pendry and A.~MacKinnon, Phys.\ Rev.\ Lett.\ {\bf 69}, 2772 (1992);
    J.~B.~Pendry, J.~Mod.\ Opt.~{\bf 41}, 209 (1994); P.~Marko\u{s} and C.~M.~Soukoulis, Phys.\ Rev.~E {\bf 65}, 036622 (2002); P.~Yeh, {\em Optical Waves in Layered Media}, Wiley, Hoboken, N.J., 2005.

\bibitem{CMP} C.~C.~Wan, T.~De~Jesus, and H.~Guo, Phys.\ Rev.~B {\bf 57}, 11907 (1998); D.~R.~Smith, S.~Schultz, P.~Marko\u{s} and C.~M.~Soukoulis, Phys.\ Rev.~B {\bf 65}, 195104 (2002); P.~Marko\u{s}, Acta Phys.\ Slovaca~{\bf 56}, 561 (2006).

\bibitem{Accu} W.~T.~Thompson, J.~Appl.\ Phys.~{\bf 21}, 89 (1950);  D.~Levesque and L.~Piche, J. Accous.~Soc.\ Am.~{\bf 92}, 452 (1992);  B.~Hosten and M.~Castaings, J. Accous.~Soc.\ Am.~{\bf 94}, 1488 (1993).
    
\bibitem{invisible-0} L.~Poladian, Phys.\ Rev.\ E~{\bf 54}, 2963 (1996); M.~Greenberg and M.~Orenstein, Opt.\ Lett.~{\bf 29}, 451 (2004); M.~Kulishov, J.~M.~Laniel, N.~Belanger, J.~Azana, and D.~V.~Plant, Opt.\ Exp.~{\bf 13}, 3068 (2005).

\bibitem{invisible} Z.\ Lin, H.\ Ramezani, T.\ Eichelkraut, T.\ Kottos, H.\ Cao, and
D.\ N.\ Christodoulides, Phys.\ Rev.\ Lett.\ {\bf 106}, 213901 (2011). See also \cite{invisible-0}.

\bibitem{invisible-other} E.~M.~Graefe and H.~F.~Jones, Phys.\ Rev.~A {\bf 84}, 013818 (2011);
S.~Longhi, J.~Phys.~A {\bf 44}, 485302 (2011); H.~F.~Jones, J.~Phys.~A {\bf 45}, 135306 (2012).

\bibitem{exp} A.~Regensburger, C.~Bersch, M.~A.~Miri, G.~Onishchukov, D.~N.~Christodoulides,  and U.~Peschel, Nature  {\bf 488}, 167 (2012).

\bibitem{note2} The pseudo-adjoint \cite{p1} of $\sH(\tau)$, which is defined by $\sH(\tau)^\sharp:=\sigma_3^{-1}\sH(\tau)^\dagger\sigma_3$ with $\sigma_3$ being the diagonal Pauli matrix, satisfies $\sH(\tau)^\sharp=\left[\frac{v(\tau/k)^*}{v(\tau/k)}\right]\sH(\tau)$. Therefore, if $v$ is real-valued, $\sH(\tau)$ is pseudo-Hermitian \cite{p1}, and if $v$ is complex-valued, we have $[\sH(\tau)^\sharp,\sH(\tau)]=0$, i.e., $\sH(\tau)$ is pseudo-normal.
    
\bibitem{muga} J.\ G.\ Muga, J.\ P.\ Palao, B.\ Navarro, and I.\ L.\ Egusquiza, Phys.\
Rep.\ {\bf 395}, 357 (2004).

\bibitem{IST} K.~Chadan and P.~C.~Sabatier, {\em Inverse Problems in Quantum Scattering Theory,} Springer, Berlin, 1989.

\bibitem{p1} A.~Mostafazadeh, J.\ Math.\ Phys.~{\bf 43}, 205, 2814, and 3944 (2002); Int.\ J.~Geom.\ Meth.\ Mod.\ Phys.~\textbf{7}, 1191 (2010).

\bibitem{muga-1991} J.\ G.\ Muga, Phys.\ Lett.~A \textbf{157}, 325 (1991).



\end{thebibliography}
